\documentclass[a4paper,11pt]{article}
\usepackage{pos}

\newcommand{\al}{a_{\rm L}}
\newcommand{\bl}{b_{\rm L}}
\newcommand{\kt}{k_{\rm T}}
\newcommand{\ncoll}{n_{\rm coll}}
\newcommand{\pt}{p_{\rm T}}
\newcommand{\raa}{R_{\rm AA}}
\newcommand{\rpa}{R_{\rm pA}}
\newcommand{\rppb}{R_{\rm pPb}}
\newcommand{\siglq}{\sigma_{\rm LQ}}
\newcommand{\sighq}{\sigma_{\rm HQ}}

\title{Resolving the $R_{\rm pA}$ and $v_2$ puzzle of $D^0$ mesons in $p-$Pb collisions}

\author[a,b]{Chao Zhang}
\author*[b]{Zi-Wei Lin}
\author[c]{Liang Zheng}
\author[d]{Shusu Shi}
\affiliation[a]{
School of Science, Wuhan University of Technology, Wuhan, 430070, China}
\affiliation[b]{
Department of Physics, East Carolina University,Greenville, North
Carolina 27858, USA}
\affiliation[c]{China University of Geosciences, Wuhan 430074, China}
\affiliation[d]{Key Laboratory of Quark  Lepton Physics (MOE) and
  Institute of Particle Physics, Central China Normal University,
  Wuhan 430079, China}
\emailAdd{linz@ecu.edu}

\abstract{
It has been difficult to reconcile the experimental data on the $D^0$
meson nuclear modification factor and elliptic flow in $p-$Pb
collisions at LHC energies. Here we study these observables with the
string melting version of a multi-phase transport model, which has
been improved with the implementation of the Cronin effect (or
transverse momentum broadening) and independent fragmentation for
charm quarks. Using a strong Cronin effect allows us to provide the first
simultaneous description of the $D^0$ meson $R_{\rm pA}$ and $v_2$ data at
$p_{\rm T} \leq$ 8 GeV$/c$. The model also provides a reasonable description
of the $D^0$ meson $p_{\rm T}$ spectra and the low-$p_{\rm T}$ (below $\sim$ 2
GeV$/c$) charged hadron spectra in $p+p$ and $p-$Pb
collisions as well as $R_{\rm pA}$ and $v_2$ in $p-$Pb collisions. We find
that both parton scatterings and the Cronin effect 
are important for the $D^0$ meson $R_{\rm pA}$, while parton scatterings are
mostly responsible for the $D^0$ meson $v_2$. Our results indicate
that it is crucial to include the Cronin effect for the simultaneous
description of the $D^0$ meson $R_{\rm pA}$ and $v_2$. Since the Cronin
effect is expected to grow with the system size, this work 
implies that the Cronin effect could also be important for heavy hadrons in
large systems.}

\FullConference{HardProbes2023\\
 26-31 March 2023\\
 Aschaffenburg, Germany\\}

\begin{document}
\maketitle

\section{Introduction}
The nuclear modification factors including $\raa$ and $\rpa$ and elliptic flow
$v_2$ are frequently utilized to study the hot and dense matter
generated during the collision of two nuclei at high energies. 
Recent measurements from the experiments at the LHC have revealed 
a relatively flat $\rpa$~\cite{ALICE:2019fhe} close to unity but a
significant $v_2$~\cite{CMS:2018loe} for $D^0$ mesons in $p-$Pb
collisions, which has posed a substantial challenge for the theoretical
understanding. Both hydrodynamics-based models and parton/hadron  
transport models suggest that significant interactions between charm
quarks and the quark-gluon plasma (QGP) medium
is necessary to generate a substantial $v_2$. On the other hand, 
significant interactions between charm quarks and the QGP would
inevitably suppress high-$\pt$ charm hadrons. 
Therefore, it has been difficult to explain and understand 
simultaneously the $D^0$ meson $\rpa$ and $v_2$ data. 

Several theoretical studies have successfully reproduced either 
the heavy meson $\rpa$ or the heavy meson $v_2$. For example, the
POWLANG model~\cite{Beraudo:2015wsd} can successfully describe the
$\rpa$ of heavy flavors, but it predicts a small charm
$v_2$. Perturbative QCD (pQCD) calculations that incorporate the cold nuclear matter effect can generally explain the data
on charm $\rpa$~\cite{Kang:2014hha}, as another pQCD model that
includes a parameterized $\kt$ broadening. Regarding the elliptic
flow of heavy flavors, the color glass condensate framework can
reproduce the open and hidden charm meson $v_2$ in $p-$Pb collisions
at the LHC~\cite{Zhang:2020ayy}, suggesting the significance  
of initial state correlations for heavy quarks in small
systems. However, a simultaneous description of both $\rpa$ and $v_2$ 
of heavy hadrons has not yet been achieved. Here we examine the $\rpa$
and $v_2$ of $D^0$ mesons in $p-$Pb collisions at LHC energies using
an improved version of a multi-phase transport (AMPT) model.

\section{The improved AMPT model for this study}
The AMPT model~\cite{Lin:2004en} is a Monte Carlo event generator that
contains both partonic and hadronic phases in high energy
heavy ion collisions.
The string melting version, which we use for this study, mainly
contains four parts: the fluctuated initial conditions, partonic
scatterings, quark coalescence, and hadronic scatterings. 
Recently, we have improved the AMPT model with a new quark
coalescence~\cite{He:2017tla}, incorporated modern parton
distribution functions of the free proton and a spatially-dependent
nuclear shadowing~\cite{Zhang:2019utb}, 
improved heavy flavor productions~\cite{Zheng:2019alz}, and applied
local nuclear scaling of the two key input parameters for
self-consistent dependence on the system size or
centrality~\cite{Zhang:2021vvp}. The AMPT model used in this
study~\cite{Zhang:2022fum} includes the above improvements. 
Regarding heavy quarks in the AMPT model, we also made further
improvements by isolating the initial heavy quarks from the string
melting process, including the initial state Cronin effect, and implementing 
the independent fragmentation process. 

Since the initial charm quarks are produced from hard scatterings, not
from string fragmentation, the string melting process of the AMPT
model should not apply to them. 
In this work, we separate the initial state charm quarks produced from
the HIJING model so that they enter the parton cascade (without going
through the string melting process) after a formation time
$\tau_f=E/m_T^2$, where $E$ and $m_T$ represent the charm quark energy
and transverse mass, respectively. 
We also distinguish the cross section among light quarks ($\siglq$)
from the cross section between a heavy quark and other quarks
($\sighq$). Unless specified otherwise, the default values of 
$\siglq=0.5$ mb and $\sighq=1.5$ mb are used; these values are
determined by fitting the charged hadron $v_2$ data in $p-$Pb  
collisions at 5.02 TeV and the $D^0$ meson $v_2$ data in $p-$Pb
collisions at 8.16 TeV, respectively. 
Furthermore, in addition to the usual quark coalescence process, we
have included the independent fragmentation as another 
hadronization channel for heavy quarks. When a heavy 
quark and its coalescing partner(s) have a large relative distance or
a large invariant mass, they are deemed unsuitable for quark
coalescence. In such cases, the heavy quark undergoes hadronization
through the independent fragmentation~\cite{Zhang:2022fum}. 

In addition, we include the transverse momentum broadening, known as
the Cronin effect, for the heavy quarks in the initial state (before
they enter the parton cascade). The broadening is implemented by introducing a transverse momentum kick $\kt$ to each $c\bar{c}$ pair in the initial state. The value of $\kt$ is randomly sampled from a two-dimensional Gaussian distribution characterized by a Gaussian width parameter $w$:
\begin{eqnarray}
&&f \left (\vec \kt \right )=\frac{1}{\pi w^2} e^{-\kt^2/w^2},\\
\label{fkt}
&&w=w_0 \sqrt{1+(\ncoll-i) \delta}.
\label{width}
\end{eqnarray}
In the above, $i=1$ if the $c\bar c$ pair is produced from the radiation
of one participant nucleon, $i=2$ if the $c\bar c$ pair is produced from
the collision between one participant nucleon from the projectile and
another from the target, while $\ncoll$ is the number of primary NN
collisions of the participant nucleon for the former case and the sum
of the numbers of primary NN collisions of both participant nucleons
for the latter case. 
This way, $w=w_0$ for $p{+}p$ collisions. 
For $w_0$, we take the following parameterization~\cite{Zhang:2022fum}
so that it depends on the Lund string fragmentation parameters $\al$
and $\bl$ through the
effective string tension~\cite{Lin:2004en}:
\begin{equation}
w_0=(0.35{\rm~GeV}/c)~\sqrt {\frac {\bl^0(2+\al^0)}{\bl (2+\al)}}.
\end{equation}
Details about the values of the Lund string fragmentation parameters 
in the above parameterization can be found in
Ref.~\cite{Zhang:2022fum}. 
As a result, for $p{+}p$ collisions, $w=0.375$ GeV$/c$. 
We note that the pQCD-based HVQMNR code~\cite{Vogt:2021vsc} also
employs a similar approach to implement the Cronin effect for charm
quarks. While we apply the broadening to each $c\bar c$ pair in the initial
state, the HVQMNR code applies it to each charm (anti)quark. 
For comparisons, we have calculated the average $\kt$ broadening to
each charm quark or each $c\bar c$ pair. 
For $p{+}p$ collisions at 5.02 TeV, the HVQMNR code yields $\langle 
\kt^2 \rangle = 1.46 \rm{~GeV}^2$ and $2.92 \rm{~GeV}^2$ for a single
charm quark and a $c\bar c$ pair, respectively. These values are higher
than our corresponding values of $0.04 {\rm~GeV}^2$ and $0.14
{\rm~GeV}^2$. In the case of minimum bias $p$-Pb collisions at 5.02
TeV, the HVQMNR code gives $\langle \kt^2 \rangle=2.49{\rm~GeV}^2$ and
$4.97 \rm{~GeV}^2$ for a single charm quark and a $c\bar c$ pair,
respectively, which are lower than our corresponding values of
$3.27{\rm~GeV}^2$ and $13.0{\rm~GeV}^2$ (for $\delta=5$).

\section{Results}

\begin{figure*}
    \begin{minipage}[t]{0.5\linewidth}
    \centering
    \includegraphics[scale=0.35]{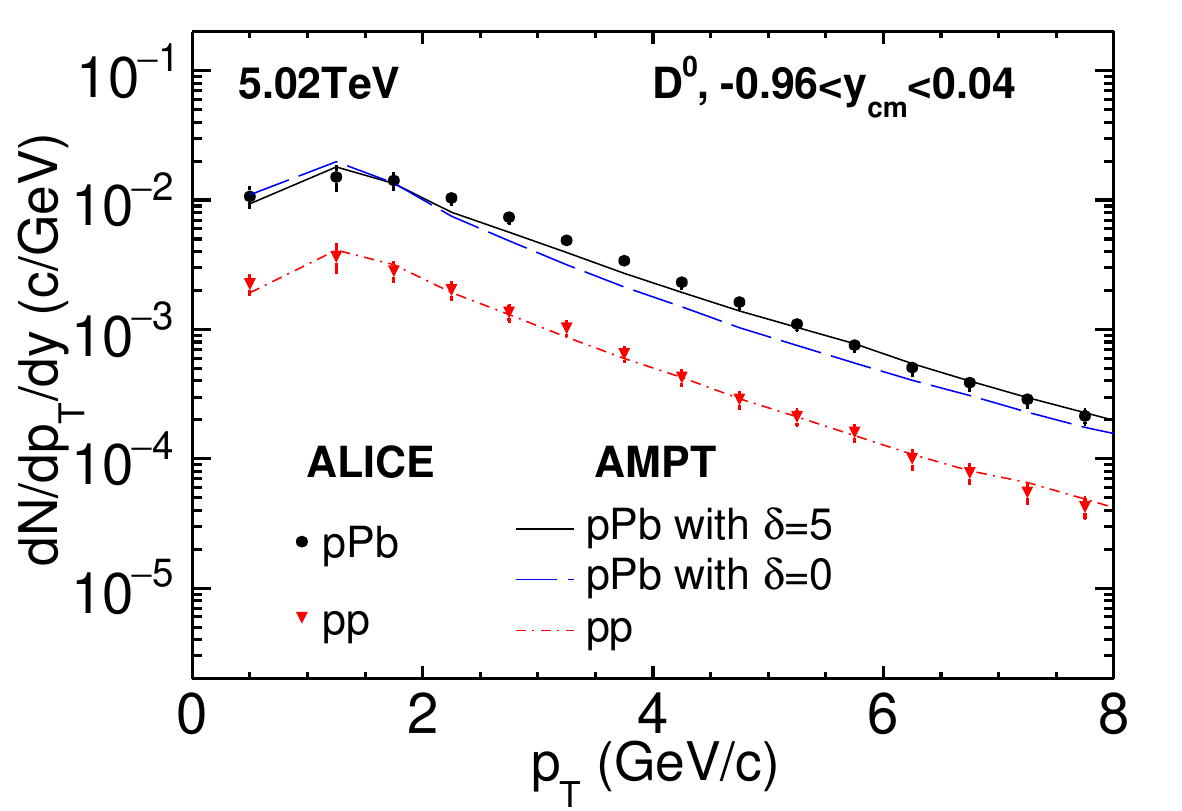}    
  \end{minipage}%
  \begin{minipage}[t]{0.5\linewidth}
    \centering
    \includegraphics[scale=0.35]{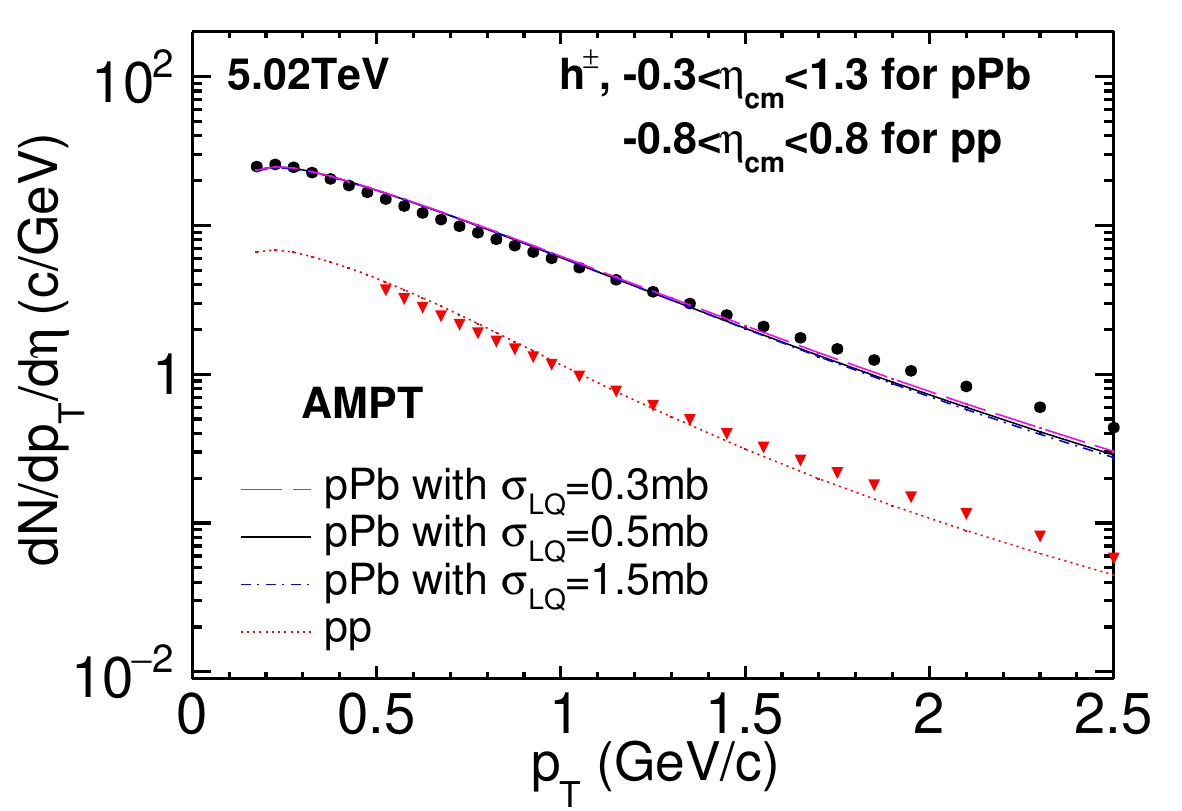}
     \end{minipage}
    \caption{Left panel: AMPT model results on the transverse momentum
      spectra of $D^0$ mesons at mid-rapidity in $p+p$ collisions and
      minimum bias $p-$Pb collisions (with $\delta=5$ or $0$) at 5.02
      TeV in comparison with the experimental data. Right panel:
      transverse momentum spectra of charged 
      hadrons around mid-rapidity in $p+p$ collisions and $p-$Pb
      collisions at 5.02 TeV from the AMPT model in comparison with
      the experimental data. The AMPT spectra for $p-$Pb collisions
      are shown for three values of $\siglq$: 0.3, 0.5, and 1.5 mb.}
     \label{fig:1}
  \end{figure*}

The left and right panels of Fig.~\ref{fig:1} show the transverse
momentum spectra of $D^0$ mesons and charged hadrons, respectively, in
$p+p$ and minimum bias $p-$Pb collisions at 5.02 TeV from the AMPT
model in comparison with the experimental data. 
The improved AMPT model provides a good description of the
experimental data for both collision systems. 
We find that the Cronin effect is very important for the $D^0$ meson
$\pt$ spectra in $p-$Pb collisions, where the AMPT model without the
Cronin effect (i.e., at $\delta=0$) underestimates the yield of $D^0$ meson
at relatively high $\pt$, while using $\delta=5$ leads to a significant
enhancement of the $D^0$ meson yield at relative high $\pt$. 
The effect of parton scatterings on the charged hadron $\pt$ spectra is
also investigated. We see that the parton scatterings will suppress the
charged hadron yield at relative high $\pt$ due to the parton energy
loss or jet quenching.

\begin{figure*}
    \begin{minipage}[t]{0.5\linewidth}
    \centering
    \includegraphics[scale=0.35]{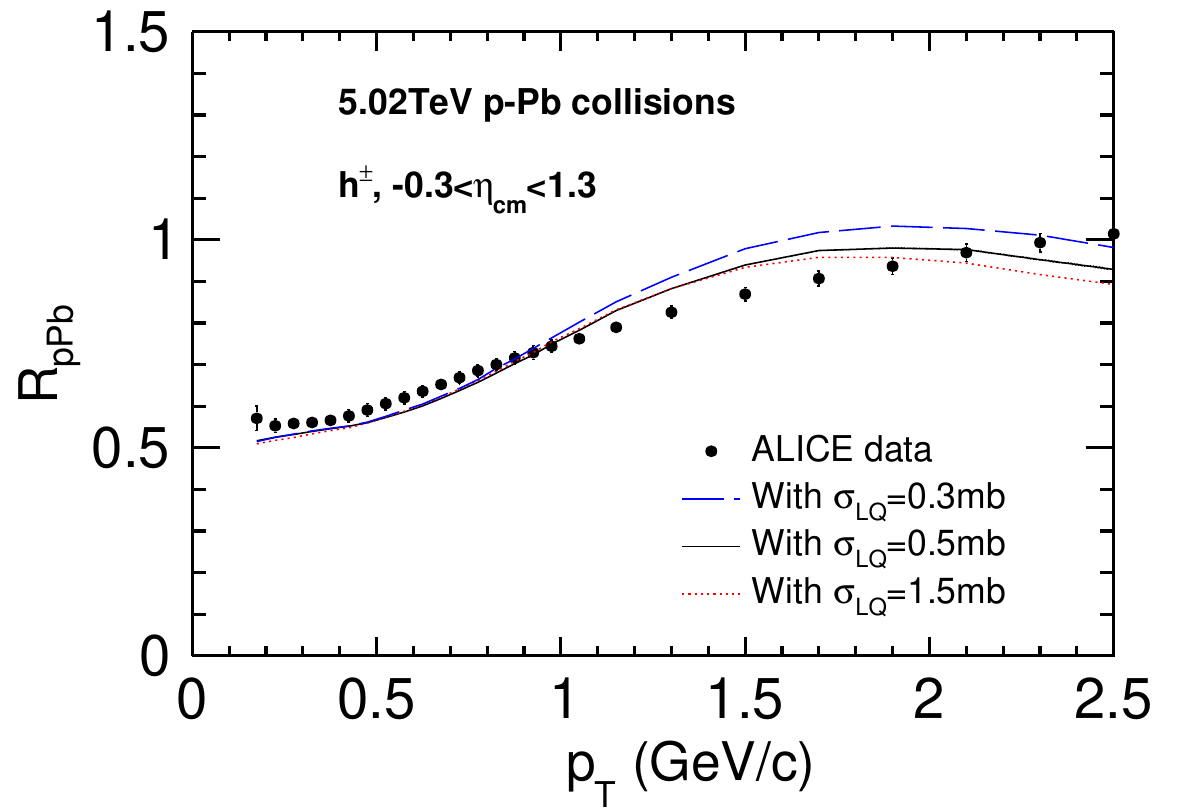}    
  \end{minipage}%
  \begin{minipage}[t]{0.5\linewidth}
    \centering
    \includegraphics[scale=0.35]{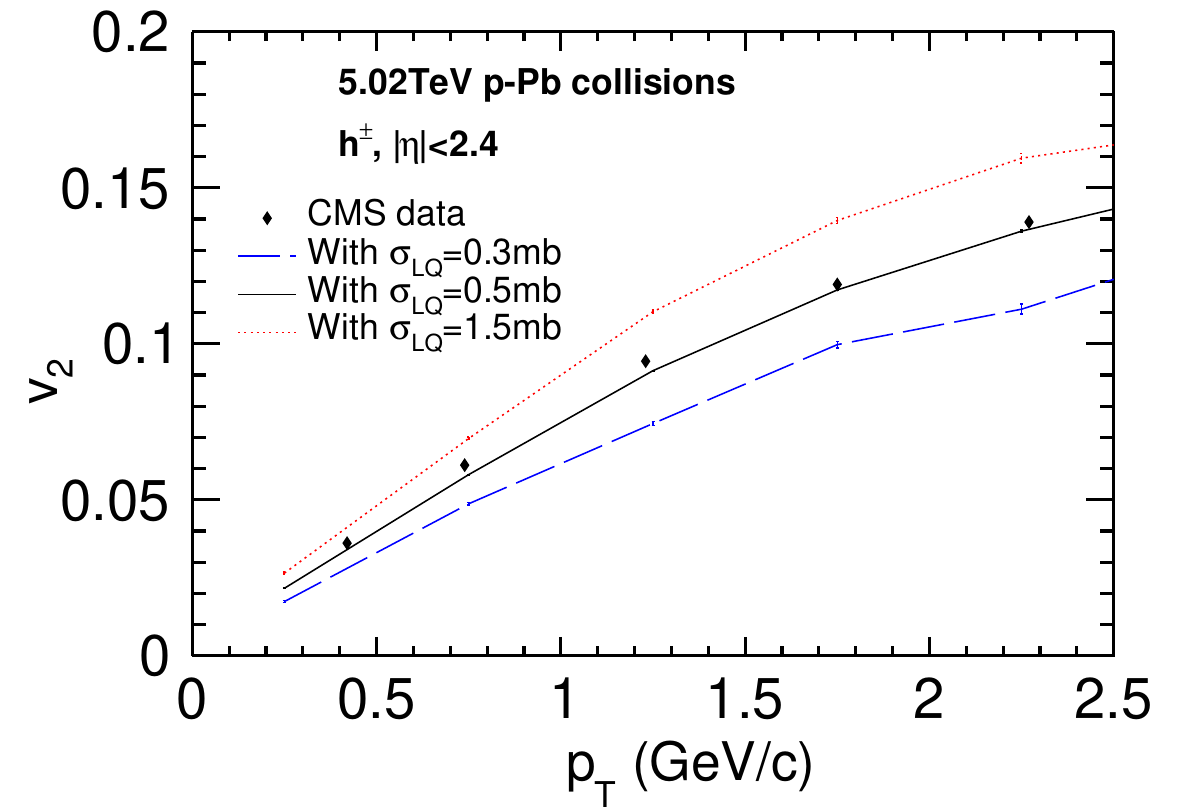}
     \end{minipage}
    \caption{Charged hadron $\rppb$ (left panel) in minimum bias
      $p-$Pb collisions at 5.02 TeV and $v_2$ (right panel) in high
      multiplicity $p-$Pb collisions at 5.02 TeV from the AMPT model 
      with $\siglq=$ 0.3, 0.5, and 1.5 mb in comparison with
      the experimental data.} 
     \label{fig:2}
  \end{figure*}

In Fig.~\ref{fig:2} we investigate the nuclear modification factor
$\rppb$ and elliptic flow $v_2$ of charged hadrons around mid-rapidity
in $p-$Pb collisions at 5.02 TeV. The AMPT model can reasonably
describe these observables up to $\pt \sim$ 2 GeV$/c$. Consistent with
Fig.~\ref{fig:1}, we see that a larger parton cross section results in
a lower (or a moderate reduction of) nuclear modification factor
$\rppb$ for charged hadrons at relatively high $\pt$. 
On the other hand, a larger parton cross section leads to a
substantial increase in the elliptic flow ($v_2$) of charged hadrons. 

\begin{figure}[!htb]
\includegraphics[scale=0.75]{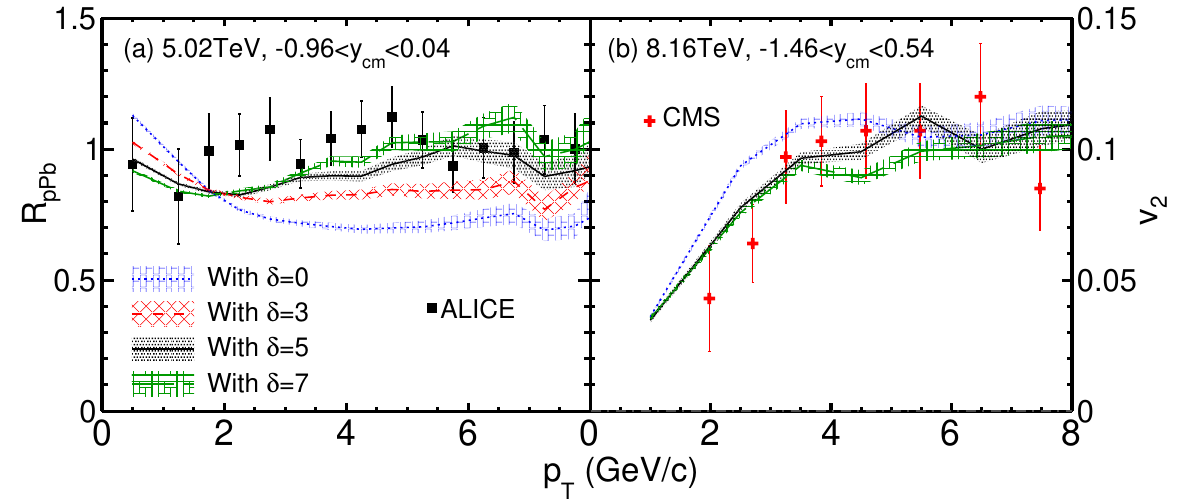}
\caption{$D^0$ meson (a) $\rppb$ in minimum bias  $p-$Pb collisions at
  5.02 TeV and (b) $v_2$ in high multiplicity $p-$Pb collisions at
  8.16 TeV from the AMPT model with different strengths of the Cronin
  effect ($\delta$) in comparison with the experimental data
  (symbols).} 
\label{fig:3}
\end{figure}

We now examine the nuclear modification factor ($\rppb$) of $D^0$
mesons in $p-$Pb collisions at 5.02 TeV and their elliptic flow
($v_2$)  in $p-$Pb collisions at 8.16 TeV. 
Figure~\ref{fig:3}(a) compares the $D^0$ meson $\rppb$ data 
with the model results for different strengths of the Cronin effect
(via the $\delta$ parameter). 
Figure~\ref{fig:3}(b) shows the $v_2$ of $D^0$ mesons
from the AMPT model in comparison with the data. 
We see that the AMPT model with a strong Cronin effect (at $\delta=5$
or $\delta=7$) provides a reasonable description of both
the $\rpa$ and $v_2$ observables. 
A large value of $\delta$ or a strong Cronin effect is found to result
in a significant enhancement of the $D^0$ meson $\rppb$ at relatively
high $\pt$,  while it leads to a small decrease of the $D^0$ meson $v_2$. 

\section{Summary}
We have studied the nuclear modification factor $\rppb$ of $D^0$
mesons and charged hadrons in minimum bias $p-$Pb collisions as well
as the elliptic flows $v_2$ in high multiplicity $p-$Pb collisions at
LHC energies with a multi-phase transport model. The model has been 
improved with the inclusion of transverse momentum broadening (i.e., 
the Cronin effect) and independent fragmentation for charm quarks. 
When invoking a strong Cronin effect, we are able to provide the first
simultaneous description of both the  $\rppb$ and $v_2$ data of $D^0$
mesons below the transverse momentum of 8 GeV$/c$. 
Our results show that both parton scatterings and the Cronin effect
significantly affect the $\rppb$ of $D^0$ mesons. On the other
hand, the $v_2$ of $D^0$ mesons is primarily generated by parton
scatterings, while the Cronin effect leads to a modest reduction of
the charm $v_2$. In particular, we demonstrate that the Cronin effect
could resolve the $\rppb$ and $v_2$ puzzle of $D^0$ mesons at LHC
energies.

\end{document}